\documentstyle[referee]{mn}
\newcommand{\reference}{\bibitem}
\newcommand{\beq}{\begin{equation}}
\newcommand{\eeq}{\end{equation}}
\newcommand{\bey}{\begin{eqnarray}}
\newcommand{\eey}{\end{eqnarray}}

\newcommand{\pc}{\,{\mbox{pc}}}
\newcommand{\mpc}{\,{\rm {Mpc}}}
\newcommand{\kpc}{\,{\rm {kpc}}}

\newcommand{\kms}{\,{\rm {km\, s^{-1}}}}
\newcommand{\msun}{M_\odot}

\newcommand{\Rd}{R_d}

\newcommand{\Md}{M_d}
\newcommand{\md}{m_d}

\newcommand{\mue}{\langle\mu\rangle_{\rm eff}}
\newcommand{\muB}{\langle\mu\rangle_{\rm eff, B}}
\newcommand{\re}{r_{\rm eff}}
\newcommand{\sigmae}{\langle\Sigma\rangle_{\rm eff}}

\newcommand{\obnow}{\Omega_{B,0}}

\newcommand{\sz}{\sigma_z}
\newcommand{\sr}{\sigma_r}
\newcommand{\st}{\sigma_\theta}
\newcommand{\rz}{\Re_z}
\newcommand{\rr}{\Re_r}
\newcommand{\rt}{\Re_\theta}
\newcommand{\mmw}{MMW}
\newcommand{\feff}{f_{\rm eff}}
\newcommand{\feffS}{f_{\rm eff}}
\newcommand{\feffE}{f_{\rm eff}}
\def\arcsecf {\hbox{$.\!\!^{\prime\prime}$}}

\input epsf

\title[] 
{On the Physical Connections between Galaxies of Different Types}
\author[]
{Shude Mao and H.J. Mo
\thanks{E-mail: (smao; hom)@mpa-garching.mpg.de} \\
	Max-Planck-Institut f\"ur Astrophysik
	Karl-Schwarzschild-Strasse 1, 85748 Garching, Germany}
\date{Accepted ........
      Received .......;
      in original form .......}
\pagerange{\pageref{firstpage}--\pageref{lastpage}}
\pubyear{1997}

\begin{document}
\maketitle
\label{firstpage}

\begin{abstract}

Galaxies can be classified
in two broad sequences which are likely to reflect their formation
mechanism. The `main sequence', consisting of spirals, 
irregulars and all dwarf galaxies, is probably produced by gas
settling within dark matter haloes. We show that the sizes and
surface densities along this sequence are primarily determined 
by the distributions of the angular momentum and formation time 
of dark haloes. They
are well reproduced by current cosmogonies provided that galaxies
form late, at $z \la 2$. In this scenario, dwarf ellipticals
were small `disks' at $z\sim 1$ and become `ellipticals'
after they fall into cluster environments. The strong clustering
of dwarf ellipticals is then a natural by-product of the merging
and transformation process. The number of dwarf galaxies predicted in
a cluster such as Virgo is in good agreement with the observed
number. On the other hand, the `giant branch', consisting of giant 
ellipticals and bulges, is probably produced by the merging of disk
galaxies. Based on the observed phase-space densities of galaxies,  
we show that the main bodies of {\it all} giant ellipticals can
be produced by dissipationless mergers of high-redshift disks.
However, high-redshift disks, although denser than
present-day ones, are still not
compact enough to produce the high {\it central} phase space density
of some low-luminosity ellipticals.
Dissipation must have occurred in
the central parts of these galaxies during the merger which formed
them.
\end{abstract}


\section {INTRODUCTION}

Galaxies exhibit a wide range in 
luminosities, sizes and shapes.
Their shapes divide them into spirals, ellipticals
and irregulars, whereas their sizes and luminosities distinguish
giants from dwarfs. Dwarf galaxies (here defined as having a magnitude
${\cal M} \ga -16$), while fainter, dominate the observed
number counts of galaxies
(see Ferguson \& Binggeli 1994 for a review).  These
galaxies can be roughly divided into two morphological classes:
dwarf ellipticals (dE's) and dwarf irregulars (dI's).
The dI's share many common properties of the more luminous disk galaxies:
their profiles are roughly exponential,
they are gas rich, rotationally
supported and mostly reside in the field. Hence they are generally
believed to be the low-luminosity extension of the more luminous 
late-type galaxies.
By comparison, dE's are more puzzling: on the one hand, they
show some similar properties to the dI's, for
example, their radial profiles are roughly exponential (Faber \& Lin 1983),
and they follow a similar correlation between the surface brightness
and the absolute luminosity (see \S 2); on the other hand, their isophotes are
elliptical, they reside mainly in clusters of galaxies and
at least some bright dE's seem to be supported by random motions
(Bender \& Nieto 1990).
 These latter properties are reminiscent of giant
ellipticals. The question naturally arises whether
these dE's are low-luminosity analogues of giant
ellipticals or have the same origin as the late-type
galaxies. So far three formation scenarios have been proposed.
The first suggests that dE's have evolved from
dI's (Searle \& Zinn 1978; Faber \& Lin 1983; Wirth \& Gallagher 1983).
The second advocates dE's as tidal debris from more massive systems 
(Gerloa, Carnevali \& Salpeter 1983; 
Hunsberger, Charlton \& Zaritsky 1996 and references therein).
 The third postulates dE's as the end-product  of
more massive galaxies which have suffered from
substantial mass loss (Vader 1986, 1987, and references
therein). In this paper, we test the first scenario in the context
of hierarchical structure formation models (White \& Rees 1978).
Specifically, we propose that
the main properties of disk galaxies, dI's and dE's are produced by gas 
settling into the center of dark matter haloes. Hence,
dE's are initially small `disks'
{\footnote{Note that we use `disks' to name objects which 
are roughly rotationally supported; whether or not 
they are flattened thin disks is not important in our discussion.}};
they are transformed into dE's in
high density regions like clusters by environmental
effects, such as pressure induced star formation  (Babul \& Rees 1992)
and/or galaxy harassment (Moore, Lake \& Katz 1997). The strong clustering
properties of the dE's are therefore inherently related to this
transformation process. We study the
range of sizes and surface brightnesses 
expected in such a scenario,
using a recent model of galactic 
disks (Mo, Mao \& White 1997, hereafter \mmw;
see also Dalcanton, Spergel \& Summers 1997). 
We take into account the
different star formation efficiency in galaxies using
a self-regulated star formation model (White \& Frenk 1991).
We show that our disk model plus a simple star formation
law can reproduce the overall observed properties quite well if the
present-day dwarf galaxies form late ($z \la 2$).
We also discuss giant ellipticals as the products
of merging of disks formed at high redshift. We find that
high-redshift disks, although denser, are still not
compact enough to produce the high {\it central} phase space density
in some ellipticals, and therefore, dissipation must have played 
some role in the final formation stage of these inner regions.
On the other hand,
the main bodies of all ellipticals {\it can}
be produced by dissipationless merging,  since their ``average''
phase space densities are comparable to those of spirals.

The outline of our paper is as follows. In Section 2, we present
the observational data to be used. In Section 3, we present
our model and compare the model predictions with the observations.
In Section 4, we study the connection
between disk and elliptical galaxies using phase space
density arguments. In Section 5, we discuss
some implications and future tests for our model.

\section{Observational Data}

In Figure 1, we show the effective radius, $\re$, and the
effective surface brightness, $\mue$, as functions of the B band
absolute magnitude, ${\cal M}_B$, for galaxies from different
sources. Here $\re$ is defined as
the radius within which half of the light is contained,
and $\mue$ is the mean surface
brightness within this radius. For all the samples,
we have adjusted the distances
and magnitudes to a Hubble constant of $h=0.5$.
The data for dwarf ellipticals (dE's) from 
Binggeli \& Cameron (1991, 1993) are shown as open circles, with
the nucleated ones shown with an additional central dot.
Their data were based on a B-band
photographic survey of dE's in the Virgo cluster. The selection
function for this sample is given roughly by the long dashed line.
The data for normal spirals are from Impey et al. (1996),
which also contain many dwarf irregulars (dI's). These data points
are shown as skeletal triangles. We have 
converted their surface brightness at the effective radius to
the effective surface brightness by subtracting 0.7 mag;
a correction appropriate for an exponential disk. The 
selection function for this sample is schematically
shown as the short dashed curve.
The data for dwarf spheroidals (dSph)
in the local group are adopted from Mateo (1997). This data set
(shown as filled squares) is in
good agreement with that in Irwin \& Hatzidimitriou (1995).
We have converted
their V band magnitudes using $B-V \approx 0.8$. The effective
surface brightness is derived from the
central surface brightness by assuming an exponential profile.
Fig. 1 also schematically shows the region populated by the
ellipticals and galactic bulges from
Burstein et al. (1997, hereafter BBFN; 
see also Bender, Burstein \& Faber 1992). To avoid overcrowding,
no data points are shown.
Note that this large data set includes all types of galaxies. Furthermore it 
not only lists $\re$, $\mue$, but also gives other kinematic
information, such as the central velocity dispersion. This kinematic
information will be used to study the phase space density for
disk and elliptical galaxies in \S 4.

\begin{figure}
\epsfysize=16cm
\centerline{\epsfbox{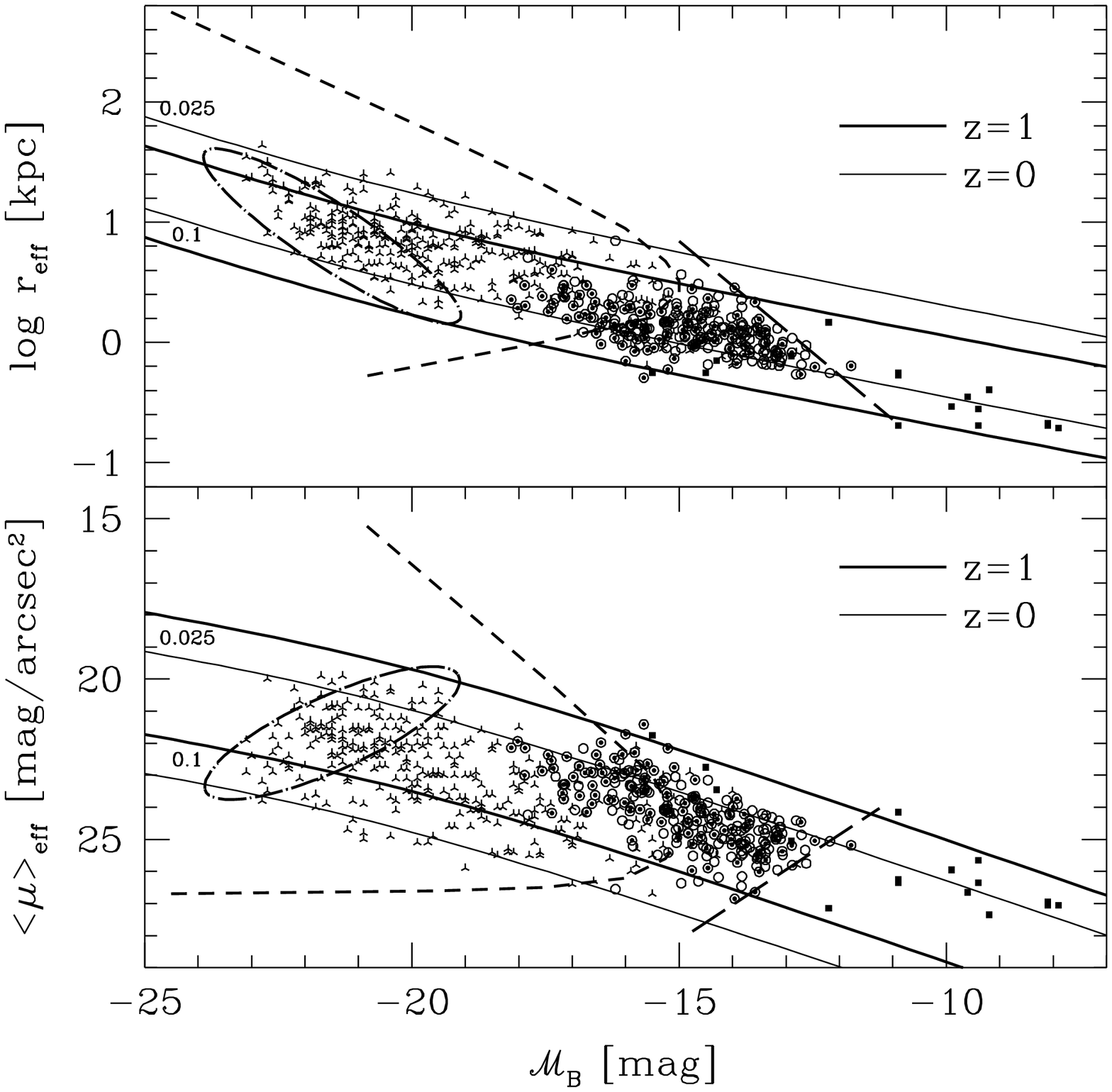}}
\caption{Effective radii, (upper panel), $\re$,
and effective surface brightnesses (lower panel),
$\mue$, are shown as a function of the B-band absolute magnitude,
${\cal M}_B$, for different types of galaxies.
The spiral and dwarf galaxies 
from Impey et al. (1996) are shown as the skeletal triangles.
The open circles are for the dE's and dS0's from Binggeli \& Cameron (1993)
while the open circles with central dots represent the
nucleated dwarf galaxies.
The filled squares indicate the dSph's 
in the local group (Mateo 1997). The short and long thick dashed lines are the
selection functions for the Impey et al. sample and the
Binggeli and Cameron sample, respectively.
The galaxies observed in these samples are to the left of these curves. 
The region populated by the elliptical galaxies and bulges 
(Burstein et al. 1997) are indicated by the dot-dashed curves.
The thin and thick solid lines are the predicted curves for
disk galaxies assembled at the present day and at $z=1$
in the SCDM model, respectively.
For each formation redshift, two curves are shown
for two spin parameters, $\lambda=0.025$ and $\lambda=0.1$.
The predictions for other structure formation models
are similar.
}
\end{figure}

As one sees from the figure, the distributions of galaxies
in the $\mue$-${\cal M}_B$ and $\re$-${\cal M}_B$ planes seem to follow
two well-defined sequences. The broad `main sequence' is defined
by the spirals, dE's, dI's and dSph's. In this sequence,
brighter galaxies have systematically higher 
effective surface brightness (i.e. lower value of $\mue$),
a trend first noticed by Kormendy (1977, see also Binggeli 
\& Cameron 1991). Note that the sequence is also clearly modulated 
by the selection effects. The continuation of the trend from dSph's, 
dI's and dE's to normal spirals, plus the fact that the profiles
of these galaxies are fairly good exponentials,
is suggestive of a common origin for these classes of objects.  
For a given absolute magnitude, the range
in $\re$ is about one decade while the scatter in $\mue$ 
is about five magnitudes. The question we want to address
here is what determines these ranges.

The other sequence, which we term as the `Giant branch', is defined by 
the giant ellipticals and galactic bulges. Unlike the main
sequence, brighter galaxies in the Giant branch have systematically
lower surface brightness, and their profiles follow more closely
the $r^{1/4}$-law. These suggest a different
formation mechanism from that of the main sequence. 
The question is then what distinguishes these two sequences and 
what are their connections.   

\section{The Main Sequence}

\subsection{The model}

 We consider galaxy formation in a cosmological context.  
For simplicity, we will only consider 
an Einstein-de Sitter universe, with $\Omega_0=1$
and no cosmological constant. The present day Hubble constant is
written as $H_0=100 h\kms\mpc^{-1}$, and we take $h=0.5$.
The initial density perturbation power spectrum
is taken to be that of the standard cold dark matter (SCDM) model with
$\sigma_8=0.6$. Our results do not change significantly
when other cosmogonies are used.

In a recent paper, Mo, Mao \& White (1997, hereafter \mmw) 
studied the formation of disk galaxies in the context
of hierarchical structure formation (e.g. White \& Rees 1978; 
White \& Frenk 1991). The model reproduces
the observed properties of disk galaxies reasonably well. Briefly, this model
assumes that the gas and dark matter are initially 
uniformly mixed. Due to dissipative and radiative processes,
the gas component adiabatically settles into a disk. Under the assumptions
that the resulting mass profile is exponential and that there is
no angular momentum loss from the gas to the dark matter,
the size and mass profile of a disk can be determined. 
 The initial density profiles of dark haloes are modeled by
\beq \label{profile}
\rho (r)={V_h ^2\over 4\pi G r^2}
{1\over \left[\ln(1+c)-c/(1+c)\right]}
{r/r_h\over (r/r_h+1/c)^2},
\eeq
where
\beq
V_h^2={GM \over r_h}, \,\,\,\,\, r_h=\left[{GM\over 100H^2(z)}\right]^{1/3},
\eeq
with $M$ being the mass of the halo, and $H(z)=H_0(1+z)^{3/2}$
the Hubble's constant at the time when the halo is assembled 
(Navarro, Frenk \& White 1996, hereafter NFW). 
The quantity $c$
in equation (\ref {profile}), called the concentration factor
of the halo, can be calculated for a given halo in any given
cosmogony (see NFW). The mass of the disk formed
in a halo at redshift $z$ is then
\beq \label{md_sis}
\Md = \md M 
\approx 1.7\times 10^{11}\msun h^{-1}
\left({\md\over 0.05}\right)\left({V_h\over 250\kms}\right)^3
\left[{H(z)\over H_0}\right]^{-1},
\eeq
where $\md$ is the fraction of the total halo mass which
settles into the disk.
We take $\md = 0.05$, consistent with the baryon density, $\obnow$,
derived from cosmic nucleosynthesis with $h=0.5$ (Walker et al. 1991). From 
the (assumed) conservation of angular momentum of a disk 
during collapse, the disk scale length and 
surface density are given by (\mmw)
\beq \label{rd_sis}
\Rd ={1\over \sqrt{2}}\lambda r_h F_R
\approx 8.8 h^{-1}\kpc \left({\lambda \over 0.05}\right)
\left({V_h \over 250\kms}\right)
\left[{H(z)\over H_0}\right]^{-1} F_R,
\eeq
\beq \label{sig_sis}
\Sigma_0 = {M_d \over 2 \pi \Rd^2} \approx 380 {M_\odot \over {\rm pc}^2}
h\left({\md \over 0.05}\right)
\left({\lambda\over 0.05}\right)^{-2}
\left({V_h \over 250\kms}\right)
\left[{H(z)\over H_0}\right] F_R^{-2},
\eeq
where $\lambda$ is the spin parameter of the halo, and the 
factor $F_R$ takes into account the disk self-gravity and
can be approximated as
\beq \label{F_R}
F_R \approx
\left({\lambda\over 0.1}\right)^{-0.06+2.71\md+0.0047/\lambda}
(1-3\md+5.2 \md^2) { 1-0.019 c+0.00025 c^2+0.52/c \over
[{2/3}+({c/21.5})^{0.7}]^{1/2} }
\eeq
(see \mmw~ for details).

 Equations (\ref{profile})-(\ref{F_R}) determine the mass
distribution in our model.
To make a connection to the observed light distribution, we need
to specify how efficiently the baryonic gas forms stars.
Unfortunately, this process is not well understood; so far only
phenomenological descriptions of star formation have been developed.
White \& Frenk (1991) argued that star formation is self-regulated,
and the star formation efficiency in a galaxy 
is determined by the balance of supernovae heating
and radiative cooling. This model predicts a star formation efficiency 
that is correlated with halo circular velocity:
\beq \label{eps0}
\epsilon_* = {1 \over 1+ \epsilon_0 (700 \kms/V_h)^2}.
\eeq
For $\obnow=0.05$, White \& Frenk 
(1991) found that $\epsilon_0=0.03$ is compatible with
the observed luminosity density in the universe and the 
mean metallicity of galaxies. This is similar to the
value $\epsilon_0=0.02$ derived by Dekel \& Silk (1986) based
on a more elaborate treatment of supernova
remnant evolution. We shall take $\epsilon_0=0.03$,
while cautioning that this parameter is
rather uncertain (cf. White \& Frenk 1991). 
To calculate the expected light from the mass that can form stars, 
we also need to adopt a {\it stellar\,} mass-to-light ratio. 
We assume this mass-to-light ratio to be universal 
in different galaxies and across the surfaces of disks. This
is clearly an oversimplification, since the mass-to-light
ratio depends on the detailed star formation and merging histories.
However, this is probably sufficient for
our purpose since we only concentrate on the general trend of galaxies.
We shall take a nominal value of $\Upsilon_B=2.4 h$
derived from dynamical studies of disks (Bottema 1997).

 From our assumption that disks have
exponential surface brightness profiles, the effective
radius and the effective surface density can be obtained as
\beq \label{r-sigma-eff}
\re \approx 1.67 \Rd, ~~
\sigmae \approx 0.36 \Sigma_0.
\eeq
Combining equations (\ref{rd_sis}-\ref{r-sigma-eff}) yields the
effective surface brightness 
\beq
\muB = 23.1 - 2.5 \log \left({\Sigma_0\over  10^2 M_\odot \pc^{-2}} 
{1\over \Upsilon_B} {\epsilon_*} \right),
\eeq
where the central surface density is expressed in units of
$M_\odot\pc^{-2}$. From equation (\ref{md_sis}),
the absolute magnitude of the galaxy is given by
\beq
{\cal M}_B = -19.5 - 2.5 
\log \left( {M_d \over 10^{10} M_\odot} {1 \over \Upsilon_B}
{\epsilon_*} \right).
\eeq

In Figure 1, we show $\re$ and $\mue$ as functions of 
${\cal M}_B$ predicted for disks at $z=0$ (thin lines) 
and $z=1$ (thick lines).
Here $z$ refers to the redshifts at which haloes of the
disks are assembled. For each $z$, the upper and lower curves 
correspond to $\lambda=0.1$ and $\lambda=0.025$, respectively. 
These two values are approximately the upper and lower 10 percentage 
points of the $\lambda$ distribution (see \mmw). Thus, the two
curves represent roughly the upper and lower limits on
$\re$ [or $\mue$] for a given formation redshift.
 It is clear that the observed ranges
are well reproduced as a result of the distribution
of the spin parameter $\lambda$, provided $z\la 2$. 
This means that the haloes of the galaxies in the main sequence  
should have been assembled quite recently. This conclusion
for present-day disk galaxies has been reached by \mmw~ and
in earlier work (e.g., White \& Frenk 1991).
As we will show below, such late formation is also required
for dwarf galaxies, in order for them not to merge
into bigger galaxies. 

Before we move on to the next subsection, we examine the metallicity and
mass-to-light ratio of galaxies implied by our assumed star formation
efficiency. As discussed before, the star formation efficiency
(i.e., the value of $\epsilon_0$) is 
normalized primarily by observations of giant galaxies
(White \& Frenk 1991). It is therefore interesting to see
whether this normalization gives correct results for dwarf galaxies.
We estimate the metallicity by the instantaneous recycling approximation:
\beq \label{metal}
{Z \over Z_\odot} = {y \over Z_\odot} \epsilon_*
= {y \over Z_\odot} {1 \over 1 + \epsilon_0 (700\kms/V_h)^2},
\eeq
where $y \sim Z_\odot$ is the metal yield of stars
(Binney \& Tremaine 1987, p. 565). We adopt
$y=1.2 Z_\odot$ as suggested by White \& Frenk (1991). The predicted line
is shown in Figure 2, together with the data  points
for nearby dSph's (Irwin \& Hatzidimitriou 1995). Notice
that the predicted line has a scaling of $Z \propto L^{2/5}$,
since $L_B \propto M_d \epsilon_* \propto 
V_h^5$ and $Z \propto V_h^2$. This trend
is clearly in agreement with the observations. The predicted amplitude
also agrees with the observational results within a factor of two.
Irwin \& Hatzidimitriou (1995) and Mateo (1997) also derive
the ``total'' mass-to-light ratio for dSph's, based on the
assumption that mass follows light. This assumption is clearly
violated in our model, since dark matter dominates in the outer part.
A direct comparison is thus problematic. We 
approximately identify their ``total'' mass-to-light ratio as
that within the effective radius. The total
mass within the effective radius is $M_d/2+M_h(<\re)$, and
the total light is $(M_d/2) \times \epsilon_*/\Upsilon_B$. The
mass-to-light ratio is therefore given by
\beq \label{m2l}
{M \over L_B} = \Upsilon_B \left(1+ {M_h(<\re) \over M_d}
\right) {1 \over 1 + \epsilon_0 (700\kms/V_h)^2}.
\eeq
The prediction of this model is shown in the lower panel of Figure 2, where
we have taken the dark matter mass within $\re$ to equal to
the disk mass. As one can see the predicted trend of $M/L_B \propto
L_B^{-2/5}$ is in good agreement with the observations.
The amplitude is somewhat too high,
but this discrepancy should not be taken too seriously since,
as discussed above, the comparison between theory and observations
is not straightforward. In general,
the self-regulated star formation law seems to predict the right
trend for dwarf galaxies.

\begin{figure}
\epsfysize=16cm
\centerline{\epsfbox{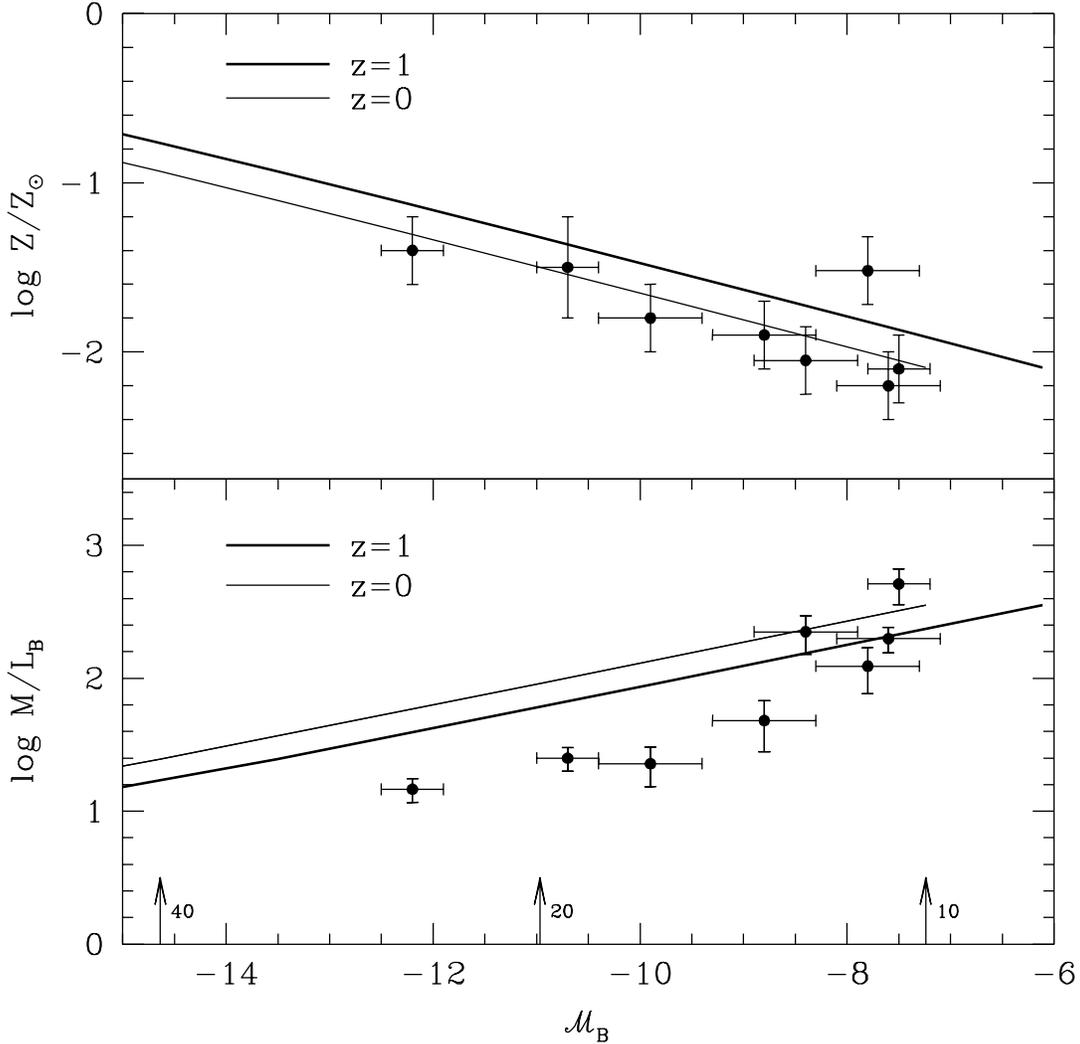}}
\caption{
Metallicity (upper panel) and mass-to-light ratio (lower panel)
are shown as a function of ${\cal M}_B$ for dSph's in the local group.
The thin and thick lines are the model predictions (cf. equations
[\ref{metal} and [\ref{m2l}]) for 
disks assembled at $z=0$ and $z=1$, respectively.
The three arrows label the magnitudes of disks 
with circular velocity of 10, 20, and 40 $\kms$ at $z=0$, respectively.
}
\end{figure}

\subsection{Connections between galaxies of different types}

As shown above, spirals, dI's, dE's and dSph's follow the same
well-defined sequence in the $\re-{\cal M}_B$ and 
$\mue-{\cal M}_B$ planes. The general trend is well reproduced by our simple
disk model. In this model, all these classes of objects start
out as `disks' forming at $z \la 2$. The disks that remain in
the field will manifest themselves as dwarf spirals and dI's.
We propose that those that have already merged 
into larger haloes such as clusters are
transformed into dwarf ellipticals at the present time (see next subsection).  
There are many processes that can perform this transformation, 
such as gas stripping by cluster potential (Faber \& Lin 1983) 
and galaxy harassment (Moore, Lake \& Katz 1997).
In the transformation process, some gas may also be compressed and flow
into the center, forming nucleated dwarf ellipticals 
(cf. Babul \& Rees 1992).
Since we link dE's directly with the cluster environment, 
they should show strong clustering, while dI's
should have the same (weak) clustering as the field spirals.
Such spatial segregations of dwarf galaxies are indeed observed
(cf. Ferguson \& Binggeli 1994).
Unfortunately, the transformation processes are difficult to quantify.
However, recent Hubble Space Telescope
(HST) observations give strong support to this picture.
These observations revealed that clusters at $z\sim 0.4$ 
contain a large number of blue late-type galaxies that have 
disappeared from clusters at the present time (Dressler et al. 1994).
These blue galaxies may just be the dwarf `disks'
formed at $z\sim 1$ which are falling into clusters 
as they are observed. Since in our model dwarf 
galaxies start out as rotation-supported objects, we expect  
late-type dwarf galaxies, such as dI's and dwarf 
spirals, to be supported by their angular momentum. 
dE's, on the other hand, may have lost some angular 
momentum in the transformation process and may no longer be 
rotationally supported. As shown by the simulations
of Moore et al. (1997), the interaction between a dwarf spiral
galaxy and cluster environment can indeed reduce the rotation substantially,
whereas the change in its effective surface density
is only modest (less than a factor of order 2).
At the moment, there are only a few dwarf elliptical galaxies
with kinematic data. The results for some low-luminosity ellipticals
suggest that bright dE's are not supported by
rotation (Bender \& Nieto 1990; see also Peterson \& Caldwell 1993).
 This is contrasted with
late-type dwarf galaxies, such as dwarf spirals and dI's, 
which are clearly supported by rotation (e.g. Salpeter \& Hoffman 1996). 
While the latter observational results are consistent
with our model prediction, a much larger kinematic sample 
is required to explore the kinematic properties of dwarf 
galaxies.

\subsection {The number and formation time of dE's
  in Virgo type clusters}

 In the model discussed above, dE's are `disks' that have merged into
larger haloes at the present time. In this section we show that this
assumption leads to specific predictions for the number density
and formation time of dwarf elliptical galaxies in clusters such as
Virgo. These predictions should be compared with 
observations, taking account of the late formation of these galaxies
discussed in \S 3.1.

 In a hierarchical clustering scenario, such as the SCDM model 
considered here, initial density perturbations are amplified by
gravitational instability, giving rise to bound clumps (dark haloes)
which grow more and more massive as they merge together and accrete
surrounding material.  Simple analytic models for such hierarchical
merging have been developed (Bond et al. 1991; Bower 1991; 
Kauffmann \& White 1993; Lacy \& Cole 1993). Using such models,
one can calculate the merging history of a dark halo. 
As a small halo merges into a larger one, its
dark matter mixes with that of the larger one to form a new halo.
However, if a galaxy (called a satellite galaxy) has already formed 
at its centre before the merger, this satellite galaxy may retain
its identity until it merges with other galaxies
in the new halo. The time scale for a satellite to merge depends
on its mass relative to that of the halo in which it is orbiting,
and can be approximated by a simple law,
\beq\label{tmerge}
t_{\rm mrg}\approx 
{1\over {\rm ln}\,\Lambda}\left({r_i\over r_h}\right)^2
\left({M_{\rm halo}\over M_{\rm sat}}\right) {r_h\over V_h},
\eeq
where $M_{\rm sat}$ is the initial mass of the satellite halo,
$M_{\rm halo}$ is the mass of the halo in which the
satellite is orbiting, $r_i$ is the initial orbital radius,
and ${\rm ln}\,\Lambda\approx 10$ is the Coulomb logarithm
(see Binney \& Tremaine 1987, \S 7.1). We assume 
$r_i=r_h/2$ to roughly take account of possible 
non-circular orbits.

 Given the merging history of a cluster and the time-scale of 
galaxy merging in dark matter haloes, one can in principle calculate
the number of satellite galaxies (cluster members) in the cluster,
and the times when the haloes of these galaxies are assembled. 
Such a treatment of galaxy formation in dark haloes has been 
developed extensively (e.g. White \& Rees 1978; White \& Frenk 1991;
Kauffmann, White \& Guiderdoni 1993; Cole et al. 1994). In this 
paper, instead of making detailed merging trees,
we define a characteristic formation time
for a (primary) halo of mass $M$ 
(identified at cosmic time $t$), $t_{1/2}$, which is the time when
the mass of its largest progenitor reaches $M/2$. 
If $t-t_{1/2}>t_{\rm mrg}$, where $t_{\rm mrg}$ is the merging time
scale (see equation [\ref{tmerge}]) for the satellite galaxy under
consideration, then the number of such satellite galaxies is given by the
number of dwarf progenitor haloes (with masses in the range
relevant to dwarf galaxies) at time $t-t_{\rm mrg}$.
This number can be easily obtained from an extension of the
Press-Schechter formalism (Press \& Schechter 1974, hereafter PS),
as discussed in Bower (1991) and Bond et al. (1991).
In this case, dwarf galaxies that fall into the cluster
earlier are destroyed by
galaxy merging, and the effective formation time of the haloes of the 
remaining dwarf galaxies is just $t-t_{\rm mrg}$. 
On the other hand, if $t-t_{1/2}<t_{\rm mrg}$, dwarf galaxies formed 
at time $t_{1/2}$ will not
be destroyed by subsequent merging of galaxies. In this case, the 
number of satellite galaxies is the sum of the number of dwarf 
progenitors haloes at time $t_{1/2}$ and the number of dwarf galaxies
contained in larger progenitors at $t_{1/2}$. The number of dwarf
progenitor haloes can, as before, be obtained from the 
extended PS formalism, while the number of dwarf galaxies 
contained in larger progenitors
can be calculated by repeating the same procedure as for the
primary halo. As a result, the total number of dwarf galaxies 
in the primary halo
can be obtained. The formation time of the halo of each dwarf galaxy
can also be identified. 

  In Figure 3, the solid curve shows the predicted numbers of dwarf 
galaxies as a function of halo circular velocity,
$V_h$. Here dwarf galaxies are defined to be those with
$B$-band luminosities $10^7L_\odot<L_B<10^9L_\odot$, to match
the observed ranges of $L_B$ for dE's (Ferguson \& Sandage
1991). This luminosity range is transformed into a mass range
of dark haloes, as described in \S 3.1. The observed numbers
of dwarf galaxies (mainly dE's) in this luminosity range
in the Virgo cluster (with a circular velocity about $800\kms$)
is about 800 (e.g. Ferguson \& Sandage 1991). This number for
small clusters like Leo and Dorado (with $V_h\sim 250\kms$)
is about 20. As one can see, the predictions are generally
in agreement with observations. Similar results have been 
obtained by Kauffmann et al. (1993) for the dwarfs 
in the local group and for the brighter galaxies (${\cal M}_B<-15$)
in the Virgo cluster. The two dashed curves show the
upper and lower quartiles of the formation redshifts of the
haloes of dwarf galaxies. It is clear that most dwarf galaxies
form rather late, at $z\la 2$, consistent with the late formation
required by the observed distributions of $\re$ and $\mue$
(see \S 3.1). The late formation of dwarf galaxies,
which is consistent with their observed intermediate age population
of stars (e.g. Held \& Mould 1994), is somewhat contrary to the 
intuition that smaller galaxies should form earlier in hierarchical 
clustering. As is clear in our model, the late formation 
is due to the fact that small galaxies that form early have
merged into larger galaxies. 

\begin{figure}
\epsfysize=16cm
\centerline{\epsfbox{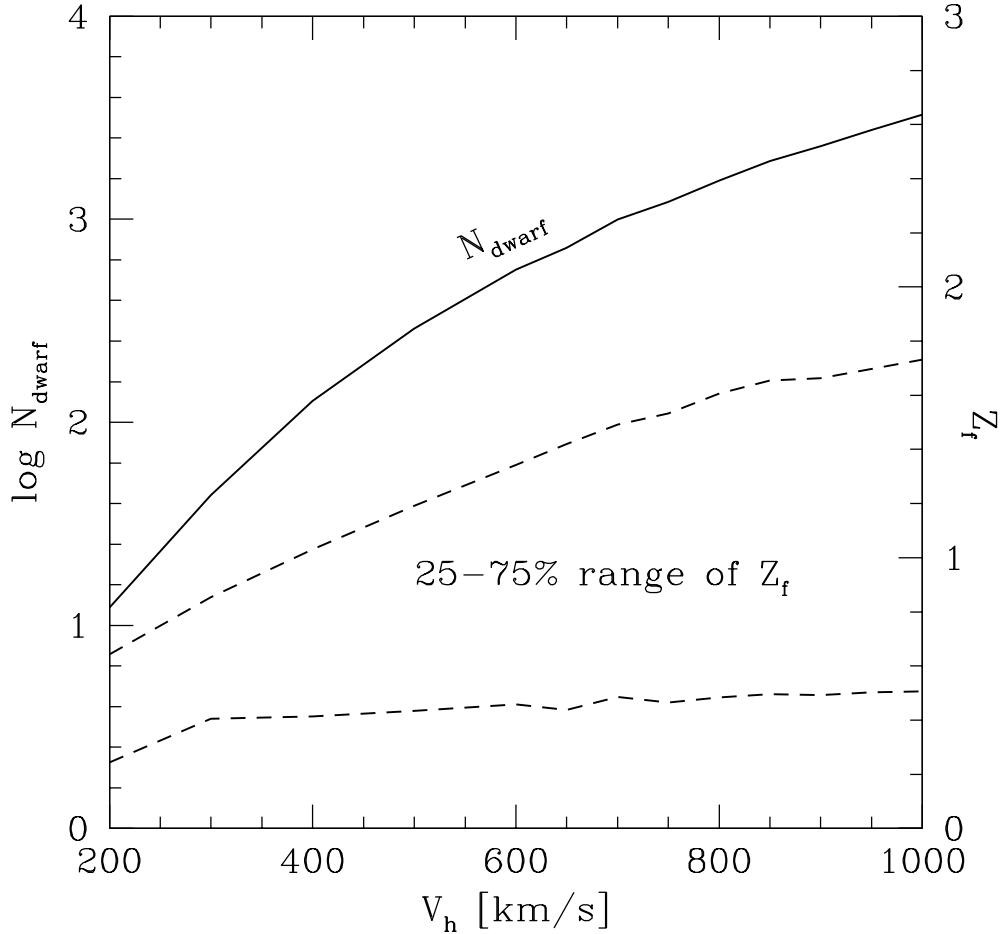}}
\caption{The solid curve shows the number of dwarf galaxies (defined in
the text) in haloes as a function of halo circular velocities.
The two dashed curves show the upper and lower quartiles of the
formation redshifts of dwarf galaxies.  
}
\end{figure}

\section{Connection to Giant Ellipticals}

Toomre \& Toomre (1972, see also Toomre 1977) proposed that 
mostly stellar disks are the building blocks from which more massive
early-type galaxies are produced. However, as noted by Carlberg
(1986; see also Hernquist et al. 1993), the central 
phase-space density in elliptical galaxies cannot be 
achieved by merging of the present-day stellar disks, since the central
phase space density in disks is lower and merging cannot 
increase the (coarse-grained) phase space density. However, there remains
two important caveats in this argument.
First, as shown in \mmw, in hierarchical clustering models,
disks formed at high redshifts are smaller and denser than present-day 
disks. Since they also form preferentially 
in high density regions and hence are more likely
to merge to form more massive galaxies, 
it is important to examine if these high redshift disks
are dense enough to be the progenitors of present-day
elliptical galaxies. Second, as pointed out by 
Hernquist et al. (1993), the maximum phase-space 
density of an elliptical may be associated only with
a small fraction of the mass at the centre of the galaxy
and so be irrelevant to the main body of the galaxy.
An important remaining question is therefore whether or not the main
bodies of elliptical galaxies can still be produced 
by stellar mergers, even though their central parts
cannot. A careful study of the phase space density of
galaxies requires detailed knowledges about stellar density 
distribution and kinematics. Unfortunately, these (particularly 
the kinematics) are not well known observationally,
therefore the calculations are only approximate. Nevertheless,
as we shall see below, the conclusions we reach are fairly robust even with
the uncertainties involved.

We first consider the central phase space density of disk and
elliptical galaxies. The central phase 
space density can be estimated by dividing the central mass density by
the volume occupied by an ellipsoid with its axes equal to
the three velocity dispersions. For ellipticals,
we adopt the same assumptions as in Carlberg (1986):
the density profile of an elliptical is a deprojected Hubble law and 
the central velocity dispersion is isothermal and isotropic.
The central phase space density is then
\beq
f_c= {27\over 16 \pi^2} {1 \over G} {1\over \sigma_c r_c^2}
= {39.5 \over \sigma_c r_c^2} {M_\odot \over \pc^3 (\kms)^3},
\eeq
where $\sigma_c$ is the central velocity dispersion in $\kms$, and
$r_c$ is the radius (in units of $\pc$) at which the surface brightness 
drops to half of the central value. For disk
galaxies, the surface mass density is radially exponential, and we model
while the vertical structure as an isothermal sheet
with a constant scale height across the disk. The central velocity dispersion
in the vertical direction is then given by 
(e.g., Binney and Tremaine 1987, p. 282)
\beq \label{sigmaZ}
\sigma_{z,0}^2 = 2 \pi G \rho_0 z_0^2, ~ \rho_0 = {\Sigma_0 \over 2 z_0},
~z_0 = \rz R_d,
\eeq
where $\rho_0$ is the central mass density, and
$z_0$ is the vertical scale height, assumed to be 
proportional to the disk scalelength by a constant factor $\rz$.
We take $\rz \approx 0.2$ (Bottema 1997).
The vertical velocity dispersion at radius $r$ can be obtained in the
same way by simply evaluating $\rho$ and $\Sigma$ at radius $r$.
We assume that the other two velocity dispersions
are proportional to $\sigma_z$:
\beq \label{couple}
\st = \rt \sz, ~~~ \sr = \rr \sz.
\eeq
The central phase space density of a disk can now be obtained
\beq\label{fc_s}
f_c = {\rho_0 \over 4 \pi\sz{_0} \st{_0} \sr{_0}/3}
={\Sigma_0 \over 2 R_d} {1 \over (\pi G \Sigma_0 R_d)^{3/2}}
{3 \over 4\pi \rz^{5/2} \rt \rr}.
\eeq
We shall take $\rt=\sqrt{2}, \rr=2$.
These values are motivated by observations of our local disk
and the flat rotation curves of disk galaxies (cf. Carlberg 1986).
Substituting eqs. (\ref{rd_sis}) and (\ref{sig_sis}) into 
equation (\ref{fc_s}) gives
\beq
f_c = 1880 h 
{\msun \over \pc^3 (\kms)^3}
{1 \over M_d}
{H(z)\over H_0}
\times \left({\lambda\over 0.05}\right)^{-3/2}
\left({\md \over 0.05}\right)^{1/2}
F_R^{-3/2}.
\eeq
Notice that for a given disk mass, the central phase space density
increases with redshift as $H(z)/H_0=(1+z)^{3/2}$.

Although the central phase space density is valuable for describing
the inner region of a galaxy,
it does not provide any direct
information for the ``global'' phase space density
in a galaxy.  We therefore need a measure
of the average phase space density of galaxies. 
Here we define such a quantity as the mass density within the
effective radius divided by the volume (in the velocity space) 
occupied by an ellipsoid with its axes equal to
the three velocity dispersions.
For disk galaxies, the mass density within the effective radius
is $\approx {\sigmae/ (2 z_0)}$, and the velocity volume is
$\approx 4 \pi \sz \sr \st/3$, therefore the effective
phase space density can be estimated as:
\beq \label{feffS}
\feffS \approx
{\sigmae \over 2 z_0} {1 \over 4 \pi \sz \sr \st/3},
\eeq
where all the velocity dispersions are evaluated at the
effective radius, $\re$. Notice that the effective phase space density 
is actually {\it higher} than the central
value because the velocity volume drops faster than the surface
density (cf. Carlberg 1986).
Similarly, one can evaluate the effective phase space density
for ellipticals, which yields the following
expression (Hernquist et al. 1993):
\beq \label{feffE}
\feffE \approx 
{15 \sqrt{3} \over 128 \pi^2} {1 \over G} {1 \over \sigma_c \re^2}.
\eeq
Equations (\ref{feffS}) and 
(\ref{feffE}) are both approximate and are accurate only within
a factor of a few.

In the top panel of Figure 4, the data points show the effective
phase space densities for the observed disks (open circles) and elliptical
galaxies (filled circles) in BBFN.
For ellipticals, $\feff$ can be obtained straightforwardly using
equation (\ref{feffE}), since
both $\sigma_c$ and $\re$ are given.
For the spirals, we first obtain the effective surface density
by using $\sigmae=\mue/(\Upsilon_B\epsilon_*)$, then
derive the velocity dispersions with
equations (\ref{sigmaZ}) and (\ref{couple}), and finally calculate
$\feff$ from equation (\ref{feffS}).
The curves in the same panel show 
the predicted phase-space densities of disks as a function of the B band 
magnitude and of the redshift $z$ when disk material is assembled.
The theoretical predictions nicely bracket the observed disk galaxies.
The effective phase space densities of spirals are higher than
those of the ellipticals because the stars are concentrated in a much 
smaller volume (i.e. in a thin plane) and because the velocity
dispersions are smaller for disks than those for the 
ellipticals. Numerical simulations show that when two
comparable stellar disks merge, the coarse-grained phase
space density drops by a large factor (Barnes 1992) and
the merger remnant resembles an elliptical galaxy. To estimate
the remnant effective phase space density, we assume that
the circular motion of the initial disk galaxy is converted
into random motions while the effective radius remains roughly the
same, as suggested by numerical simulations of disk merging 
(Barnes 1992; Weil \& Hernquist 1996).
Equation (\ref{feffE}) is then appropriate for evaluating
$\feff$ for the merger remnant.
The results are shown as open squares. They overlap 
with the data points for ellipticals, 
indicating that the main bodies of ellipticals can be 
readily produced by dissipationless mergers of disk galaxies.
If ellipticals are formed by repeated merging, 
Hernquist et al. (1993) argued that for repeated merging
of equal mass systems, the phase space density scales as 
$\propto M^{-2} \propto L_B^{-2}$. This scaling is shown as the heavy
dashed line. It reproduces the trend reasonably well, and
lends further support to the formation of ellipticals
by the merging of disk galaxies. It remains to be seen
whether the merging scenario can reproduce the 
amplitude and scatter in the observed $\re$-${\cal M}$ 
relation, and more generally the fundamental plane.

\begin{figure}
\epsfysize=16cm
\centerline{\epsfbox{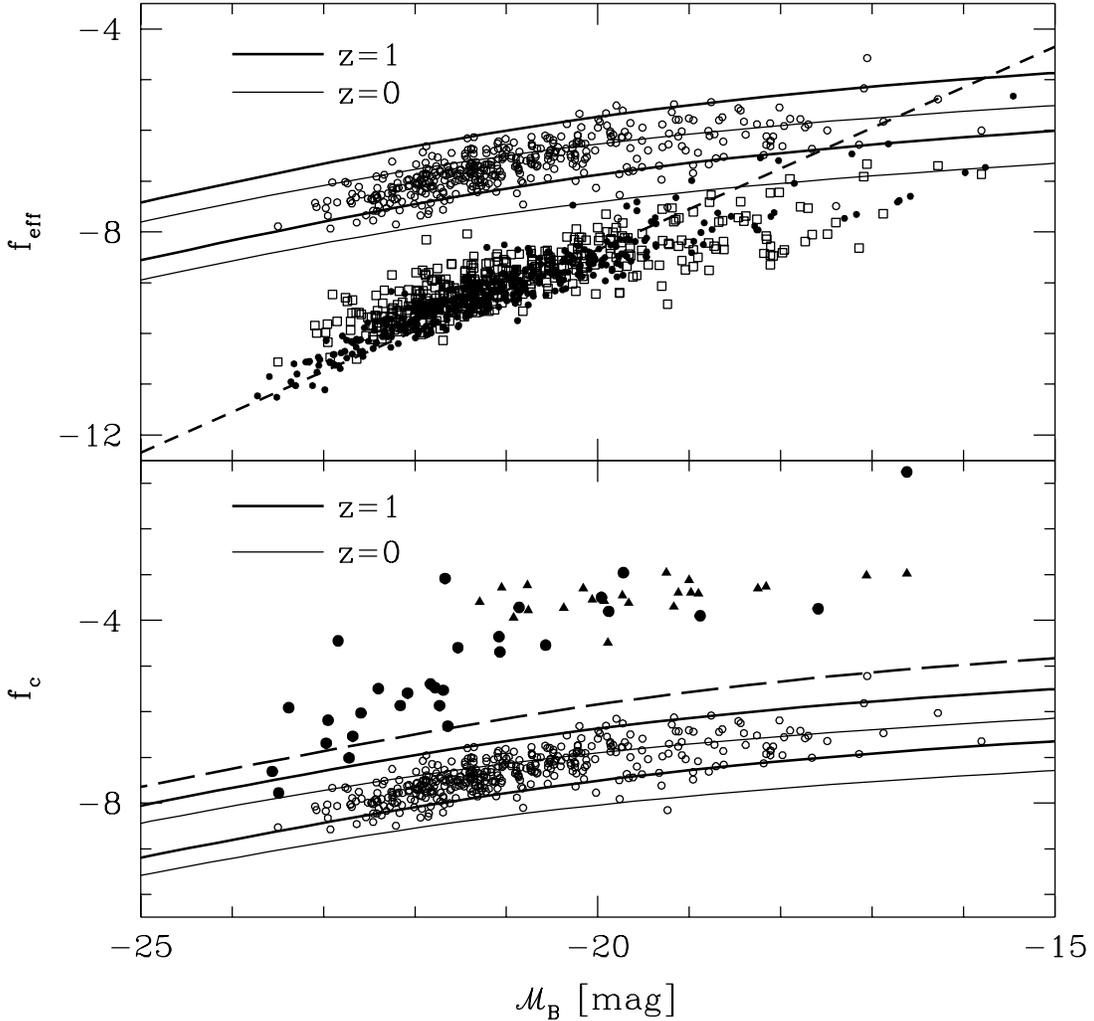}}
\caption{
The effective (top panel) and central (lower panel) 
phase space densities are shown versus the B-band absolute magnitude.
All the galaxies are from Burstein et al. (1997) except
the HST sample of Faber et al. (1997). The latter sample is used to compute the
central phase space density of ellipticals. The open circles
in the top panel indicate the effective phase space density
estimated with equation (\ref{feffS}) whereas the open squares
indicate those estimated with equation (\ref{feffE}). The
latter estimates lie on top of those for ellipticals (shown
as filled circles). The thick dashed line indicates
the slope predicted for the simplest equal-mass merging  
(Hernquist et al. 1993). In the lower panel, the filled circles
indicate the central phase space density for the core-resolved
ellipticals whereas the filled triangles indicate the lower
limits for those unresolved ellipticals. The open symbols are
again for disk galaxies.
The thick and thin solid lines are the predicted curves for
disk galaxies assembled at $z=1$ and $z=0$
in the SCDM model, respectively. 
For each formation redshift, two curves are shown
for two spin parameters,
$\lambda=0.025$ and $\lambda=0.1$. The $\lambda=0.025$ curve
is always above the $\lambda=0.1$ curve. For the central
phase space density, we have indicated one additional line
for $z=3$ and $\lambda=0.025$ (thick long dashed).
The range predicted for other structure formation models
is similar to the one shown here.
}
\end{figure}

The symbols in the lower panel of Figure 4 show the observed central
phase space densities for disks (open circles) and elliptical galaxies
(filled symbols). The data for elliptical galaxies are taken
from the Faber et al. (1997) HST sample. This sample is ideal
because it is designed to study the centres of elliptical
galaxies with the excellent HST resolution.  Even with
the excellent resolution of HST, many ellipticals
fainter than $-22$ magnitude do not show resolved cores at the HST resolution,
$0\arcsecf1$.
For these ellipticals, the half-light
radii are estimated by assuming that the central surface brightness 
is equal to that at $0\arcsecf1$.
The core radii derived in this way are obviously
upper limits and the resulting $f_c$
lower limits. These are shown as filled triangles, in contrast
with the resolved ellipticals shown as filled dots.
The data for spirals are again 
from BBFN. $f_c$ is estimated using analogous steps to
that for $\feff$ (see last paragraph). This procedure
gives the central phase space densities for
exponential disks and therefore does not take into account
the contribution from bulges, which may have higher
central phase space densities.
Despite these uncertainties, it seems clear that the
central phase space density for ellipticals fainter than $-22$ magnitude
are some three orders of magnitude higher than the observed
disks. For giant ellipticals with ${\cal M}_B \la -22$
the difference is smaller, about a factor of ten.

  The solid curves show the predicted central phase space densities
for disks at $z=0$ and 1, with $\lambda=0.025$ and 0.1.
As before, the model predicts the right range of $f_{\rm c}$
for disk galaxies. In addition, the long dashed curve
shows $f_{\rm c}$ for disks at $z=3$ with $\lambda=0.025$. 
As one can see, although the central phase space densities
are higher for disks at high redshifts, the increase is
too small to match the observed $f_{\rm c}$ for the
low-luminosity ellipticals. Thus, the central parts of 
elliptical galaxies, particularly those fainter than $\ga -22$ magnitude,
cannot be formed by dissipationless merging of
pure stellar disks. For brighter ellipticals, it is less clear
whether dissipation is necessary, since their central phase
space densities are comparable to those of fainter disks. Therefore
the central parts of the brighter ellipticals can be produced
either by dissipation or by
small disks (at high redshift) that have merged and settled into the
centres of these ellipticals.

\section{Discussion}

We have examined the connections between galaxies of different types
and the two sequences that they seem to follow:
the ``main sequence'' followed by the disk galaxies, dI's
and dE's, and the ``giant sequence'' followed by the elliptical galaxies. 
For the `main sequence', we show that angular momentum has played
a crucial role in determining the size and surface brightness
of  galaxies. For a given circular velocity,
the size of objects scales as $\lambda (1+z)^{-3/2}$, while the
surface density of object scales $\lambda^{-2} (1+z)^{-3/2}$ (cf.
equations [\ref{rd_sis}-\ref{sig_sis}]). For reasonable ranges of spin
parameter $\lambda=0.025-0.1$ and formation redshift $z=$0 to 1, these
simple scalings predict a range of size and surface brightness of
about one decade and 4.2 magnitudes, in good agreement
with the observations. These predicted ranges 
are nearly independent of detailed model parameters,
such as cosmological parameters and mass-to-light ratios.
We found that the sequence followed by elliptical galaxies and
galactic bulges can be produced
by the merging of disk galaxies. The main bodies of
ellipticals can be formed by dissipationless merging while
some dissipation must have occurred in the central parts
of some low-luminosity ellipticals.
This conclusion is based on a comparison of the central
and the ``average'' phase space densities of disk and elliptical
galaxies.

The role that the angular momentum plays in disk galaxies and
dI's is self-evident since these systems are flattened and are clearly
influenced by rotation. However, for the dE's the evidence for rotational
support is weak. In fact, a few bright dE's are known
to be supported by random motions (Bender \& Nieto 1990). Unfortunately,
the sample so far is very small and is only limited to bright dwarfs,
it remains a possibility that the fainter systems are influenced by
rotation. Even for the bright dwarfs, the velocity profiles
are only measured in the central region (with radius $\la 0.5\re$),
while in our scenario, most angular momentum is outside this
region (for an exponential disk, half of the angular
momentum is outside $2.5\re$).
It will be very interesting to extend the observational
sample to fainter dwarfs and also to the outer parts of galaxies.
In this picture, dE's are initially small disk systems, which
are later transformed into dE's, by processes such as
galaxy harassment (Moore et al. 1997) or tidal stripping
(Faber \& Lin 1983) when they enter clusters. The predicted number of dE's 
that have merged into and survived in clusters is in good
agreement with the observed numbers in clusters such as Virgo.
This conversion process is directly supported by
the HST observation of clusters at $z \sim 0.4$ (Dressler et al. 1994),
which found that bulgeless disks dominate the number
counts at the faint end. These galaxies have apparently
been transformed into dE's since these galaxies dominate nearby
clusters such as Virgo and Coma.

\section*{Acknowledgments}

We are grateful to Ralf Bender, Yipeng Jing,
Roberto Saglia, David Syer, and particularly
Simon White for helpful discussions.
We also thank Ralf Bender for providing us with data prior to publication.
This project is partly supported by
the ``Sonderforschungsbereich 375-95 f\"ur Astro-Teilchenphysik'' der
Deutschen Forschungsgemeinschaft.

{}

\newpage

\bsp
\label{lastpage}
\end{document}